\documentclass[10pt]{iopart}
\usepackage{iopams}
\usepackage{graphicx,epstopdf}
\usepackage{braket}
\usepackage{color}
\usepackage{float}
\usepackage{epsfig}
\usepackage{epstopdf}
\usepackage{times}
\usepackage{cite}

\usepackage[colorlinks=true, citecolor=blue, urlcolor=blue ]{hyperref}
\begin{document}


\title[Diverging scaling with converging multisite entanglement]{Diverging scaling with converging multisite entanglement\\
in odd and even quantum Heisenberg ladders}

\author{Sudipto Singha Roy, Himadri Shekhar Dhar, Debraj Rakshit, Aditi Sen(De), and Ujjwal Sen}
\address{Harish-Chandra Research Institute, Chhatnag Road, Jhunsi, Allahabad 211 019, India}
\ead{ujjwal@hri.res.in}

\begin{abstract}
%

We investigate finite-size scaling of genuine multisite entanglement in the ground state of quantum spin-1/2 Heisenberg ladders.
We obtain the ground states of odd- and even-legged Heisenberg ladder Hamiltonians and compute genuine multisite entanglement, the generalized geometric measure (GGM), 
which  shows that for even rungs, GGM increases for odd-legged ladder while it decreases for even ones. Interestingly, the ground state obtained by short-range dimer coverings,
under the resonating valence bond ansatz, encapsulates the qualitative features of GGM for both the ladders. We find that while the quantity converges to a single value for higher
legged odd- and even-ladders, in the asymptotic limit of a large number of rungs, the finite-size scaling exponents of the same tend to diverge. The scaling exponent of GGM
is therefore capable to distinguish the odd-even dichotomy in Heisenberg ladders, even when the corresponding multisite entanglements merge.

\end{abstract}


\submitto{\NJP}


\noindent
\section{Introduction}
The sequences of   both   odd   and   even   numbers   mathematically   reach   the   same   infinity.   The   question   is  
whether   limits   of   functions   of   odd and even   numbers   reach   the   same   limiting   function   as   the   respective  
sequences tend to infinity. 
In many-body physics, the   dichotomy   between the  physical   properties   of   odd   and   even   quantum  Heisenberg ladders   is  well known.  
For example, the seminal work in Ref.~\cite{many} shows that
the spin   gaps of odd and even Heisenberg ladders exhibit different behavior, although the values are expected to  converge to a  
single limiting value,  with an  increase in the number   of   legs of the ladder.
%
It  may   now be  asked   whether   one   can   identify   a   physical   quantity   that   would   have   different   limiting   values  
depending   on   whether    odd­-   or   even-­legged   ladders are followed to   reach   the   infinite   2D   square  
lattice. 
In this paper, we answer this question affirmatively by identifying a quantum information theoretic quantity that does the job.
%
%
%

From the perspective of quantum many-body physics, the ground state of the Heisenberg ladder \cite{many, hightclad, daggoto} is an important physical system with a rich topological order. 
The significance of these quantum spin ladders lie in their nontrivial intermediate properties between one-dimensional (1D) and two-dimensional (2D) spin lattices. For example, 
specific characteristics of Heisenberg ladders do not extrapolate trivially from the 1D Heisenberg chain to the 2D square lattice. 
This is due to the fact that odd and even Heisenberg ladders show different physical properties: Even-legged ladders are spin-gapped and have 
exponentially decaying correlation lengths while odd-legged ladders are gapless with power-law decay \cite{gopal,many,gapped,hightclad, daggoto}. 
Investigating the odd-even dichotomy and the scaling of cooperative properties in large Heisenberg ladders remains an elusive proposition, primarily due to the unavailability of suitable analytical and numerical tools. 
Quantum correlations \cite{entanglement, modi} have been used as 
a tool to detect cooperative phenomena and topological order in ground states of Heisenberg ladders \cite{sachdev,fazio,sir-mam,wen, wen2,kit,multi_others,multiparty}. In recent years, there have been studies to understand the even-odd disparity in terms of  
entropy area law  \cite{hast}, R{\'e}nyi entropy \cite{ren}, entanglement spectra \cite{njp}, etc.
However investigating the odd-even dichotomy via the scaling of cooperative multisite properties in large Heisenberg ladders remains an elusive proposition, primarily due to the unavailability of suitable analytical and numerical tools. 

In this work, we investigate the variation in ground state properties of even versus odd Heisenberg ladders by analyzing its genuine multisite entanglement.
To characterize the quantity in the ladder states, we use a computable measure, called the generalized geometric measure (GGM) \cite{ggm} (cf.\cite{geometric}). The ground state of the Heisenberg
ladder Hamiltonian is obtained using exact diagonalization algorithms \cite{lanc,tit} for moderate system size. 
We observe that the genuine multisite entanglement behaves in qualitatively different ways for the ground states of the odd- and even-legged Heisenberg ladders -- thus, detecting the odd-even dichotomy present in the system. In particular, the GGM 
increases with increasing number of ladder ``rungs'' for odd ladders while it decreases for even ladders. We subsequently observe that in terms of the behavior of GGM,  the ground states of these ladder Hamiltonians are qualitatively 
similar to
the 
ground states obtained from the RVB ansatz. 
%
%
%
Simulating the ground state of the Heisenberg ladders using RVB states allows us to analyze the finite-size scaling of genuine multisite entanglement in relatively  large spin lattices by employing recursion methods \cite{njp,rvb_more, prl}. 
We observe that although the behavior of the GGMs for odd and even RVB ladders are qualitatively different, they converge to a single value in the asymptotic limit. Therefore, for ladders with large number of rungs, as 
the number of ladder ``legs'' are increased, the odd versus even demarcation in terms of GGM vanishes. 
However, 
evaluation of
the finite size scaling exponent of multisite entanglement for large lattices
reveals
that the scaling exponents tend to \emph{diverge} for odd and even ladders, as the number of legs are increased, even though the amount of genuine multisite entanglement \emph{converges} in the asymptotic limit.

The paper is arranged as follows. In Sec. \ref{hies}, we characterize and compute the genuine multisite entanglement in ground states of the Heisenberg ladder. We introduce the RVB ladder states in Sec. \ref{dmrm} and discuss the density matrix recursion method to obtain reduced density matrices. In Sec. \ref{ggm}, we compute GGM and study its finite-size scaling in RVB ladder states. We end with a discussion in Sec. \ref{conclude}.

\section{Characterization of genuine multisite entanglement in Heisenberg ladders}
\label{hies}
The genuine multisite entanglement of a pure quantum state can be computed using the generalized geometric measure. For an $n$-party pure quantum state, $|\psi_n\rangle$, the GGM is defined as the optimized fidelity distance of the given state from the set of all states that are not genuinely multiparty entangled:
\begin{equation}
\mathcal{G}(|\psi_n\rangle)=1-\Lambda_{max}^2(|\psi_n\rangle),
\label{eq1}
\end{equation}    
where $\Lambda_{\max} (|\psi_n\rangle ) = \max | \langle \chi|\psi_n\rangle |$ with $|\chi\rangle$ being an $n$- party state not genuinely multiparty entangled. 
We note that a pure state is not genuinely multiparty entangled if it is product across any bipartition. 
For an $n$-party pure quantum state, $|\psi_n\rangle$, consisting of the parties $A_1$, $A_2$,$\ldots$, $A_n$,
Equation~(\ref{eq1}) can be shown to be equivalent to the complete form \cite{ggm}
\begin{equation}
\mathcal{G} 
=  1 - \max \{\lambda^2_{K:L} | K\cup L = \{A_i\}_{i=1}^n, K \cap  L = \emptyset\},
\label{GGM}
\end{equation}
where \(\lambda_{K:L}\) is  the maximal Schmidt coefficients in the  bipartite split \(K: L\) of \(| \psi_n \rangle\). 

{We note that all possible bipartitions, \(K: L\), of the system are considered in Eq.~(\ref{GGM}), 
with the $K$ subsystem in the above bipartition containing all possible
combinations of $A_i$, for $i = 1, 2,\ldots, n$.}
%
The computation of GGM depends on the efficient generation of arbitrary reduced density matrices across all possible bipartition of the spin system \cite{comment}. For states where the reduced density matrices can be efficiently generated, the GGM turns out to be a computable measure of genuine multisite entanglement. Additional leverage is obtained if the state is known to be symmetric {and the maximal Schmidt coefficient is known to arise from a selected subset of all possible bipartitions}. 

\begin{figure}[h]
\begin{center}
\epsfig{figure = 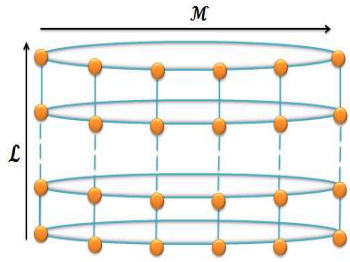, width=.36\textwidth, height=.18\textheight,angle=-0}
\caption{Schematic diagram of an $\mathcal{L}$-legged and $\mathcal{M}$-rung ladder, with $M$ (=$\mathcal{M}$) and $L$ (=$\mathcal{L}$) number of spin sites along the legs and  rungs, respectively. The boundary condition is shown by a solid line that connects the first and  last sites on a specific leg.}
\label{fig:0}
\end{center}
\end{figure}

Let us now consider the GGM of the ground state of spin-$1/2$ Heisenberg ladders, which have been
intensively studied in strongly-correlated physics in order to investigate exotic quantum phenomena, like high-$T_c$ superconductivity \cite{hightclad}, chiral Mott insulators \cite{chiral} etc.
%
%
%
%
Such studies are also interesting in view of the fact that Heisenberg models have been implemented using several experimental settings, ranging from optical lattices to nuclear magentic ressonance \cite{optical,op-ion,ion,photon, rvb_photon,nmr,rvb_optical}. 
%
The Hamiltonian of a quantum spin-1/2 Heisenberg model, with nearest-neighbor (NN) interactions, can be written as
$
H_{int}=\frac{J}{4}\sum_{|i-j|=1} \vec{\sigma_i}\cdot\vec{\sigma_{j}},
$
where $J (J>0)$ represents the NN antiferromagnetic (AFM) coupling constant. The indices, $i$ and $j$, denote the sites of an arbitrary $\mathcal{L}$-legged ladder, and  $\vec{\sigma_i}$ are the Pauli operators acting on the $i^{th}$ site. The notation $|i-j|$ indicates that the corresponding summation is over NN sites. Figure~\ref{fig:0}, shows an $\mathcal{L}$-legged ladder, with $\mathcal{M}$ rungs.

The model can not be analytically approached beyond 1D \cite{bethe-anstaz}. Though various approximate techniques such as density matrix renormalization group \cite{dmrg}, quantum Monte-Carlo \cite{qmc}, and  RVB theory \cite{rvb1,rvb2} have been used to compute certain correlation and bipartite entanglement properties, the characterization of genuine multisite entanglement in large spin systems remains an extremely challenging task. 
%
Under these restrictions, we limit our exact-diagonalization study to moderate-sized Heisenberg ladders, upto $24$ quantum spin-1/2 particles, and examine the GGM for the one-, two-, and three-legged ladders. We apply numerical algorithms, within the Lanczos method \cite{lanc, tit}, in order to obtain the ground state of the ladder Hamiltonian, and compute the GGM. 
The odd- and even-legged Heisenberg ladders show qualitatively distinct features if one considers correlation length, energy gap etc. \cite{daggoto}. We will now see whether such contrast in behavior can be seen by multipartite entanglement measure. 

\subsection{Odd-legged ladders}
\begin{figure}[h]
\begin{center}
\epsfig{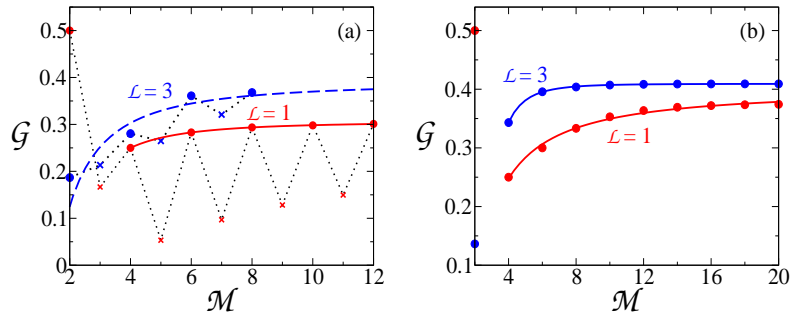}
\caption{Odd Heisenberg ladders: exact diagonalization vs. RVB ansatz. The behavior of GGM ($\mathcal{G}$), with increasing even number of rungs ($\mathcal{M}$) for odd $\mathcal{L}$-legged ladders, in ground states obtained by (a) exact diagonalization of the Heisenberg Hamiltonian and (b) using short-range RVB states. The solid lines show fits to the data values using  Eq.~(\ref{exponent}). The dashed line serves as a guide to the eye. We observe similarity in behavior between the exact and RVB ground states with respect to behavior of $\mathcal{G}$ with increasing $\mathcal{M}$. All  quantities are dimensionless.}
\label{fig:2}
\end{center}
\end{figure}

We now consider the GGM of the one- and three-legged ($\mathcal{L}$ = 1 and 3) quantum spin-1/2 Heisenberg ladders as a function of number of rungs, $\mathcal{M}$. Fig.~\ref{fig:2}(a) displays the GGM, $\mathcal{G}$, as a function of number of rungs. $\mathcal{G}$ exhibits alternating behavior based on whether the number of rungs, $\mathcal{M}$, is odd or even. This feature can not be observed in the result obtained via recursion technique of the RVB theory, described later in the text, as the RVB ansatz naturally requires an even number of rungs. From Fig.~\ref{fig:2}(a), it is evident that for both the ladders $\mathcal{G}$ increases with increasing $\mathcal{M}$. As the number of rungs increases, the rate of increment for $\mathcal{G}$ slows down rapidly. Interestingly note that the fluctuations of $\mathcal{G}$ between odd and even rungs reduces if one increases number of legs which can be observed by comparing the red lines ($\mathcal{L}$ = 1) with the blue ones ($\mathcal{L}$ = 3) in Fig.~\ref{fig:2}(a).

\subsection{Even-legged ladders}
\begin{figure}[h]
\begin{center}
\epsfig{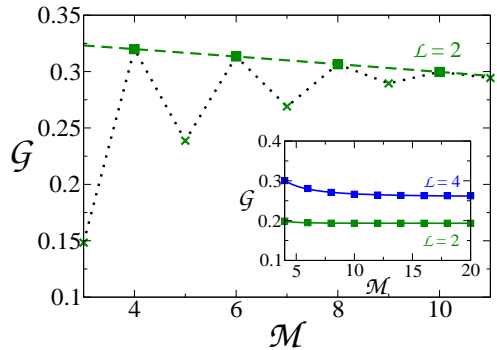}
\caption{Even Heisenberg ladders: exact diagonalization vs. RVB ansatz. The behavior of GGM ($\mathcal{G}$), with increasing number of rungs ($\mathcal{M}$) for the $2$-legged ladder, in ground states obtained by exact diagonalization of the Heisenberg Hamiltonian. The inset shows the behavior of $\mathcal{G}$ in $2$- and $4$-legged ladders using short-range RVB states. The solid lines show fits to the data values using the equation given in Eq.~(\ref{exponent}). The dashed line serves as a guide to the eye. Once again, we observe the qualitative similarity between the exact and RVB ground states with respect to the behavior of $\mathcal{G}$ with increasing even $\mathcal{M}$. All axes are dimensionless.}
\label{fig:3}
\end{center}
\end{figure}

Similarly, we also consider the GGM of the two-legged ladder ($\mathcal{L}$ = 2), as a function of number of rungs, $\mathcal{M}$ (see Fig.~{\ref{fig:3}}). We observe, $\mathcal{G}$ decreases with the increase in $\mathcal{M}$ for even rungs, while for odd rungs, it increases. However, as seen from Fig.~\ref{fig:3}, the difference of GGM between even and odd rungs decreases with the increase of rungs, and for relatively high values of $\mathcal{M}$, they both correspond to a single line following the same pattern. The same feature is obtained  for $\mathcal{L}=1$ and $\mathcal{L}=3$ in Fig. \ref{fig:2}(a).
Therefore, we conclude that with even rungs, $\mathcal{G}$ increases for odd legged ladders while decreases for even ones.

%
%




At this point, a question that arises is whether the distinct qualitative features obtained for the GGM using exact numerical simulations of the Heisenberg model can be modeled by using the RVB ansatz \cite{rvb1,rvb2}. 
This is motivated by the fact that several studies  have observed the odd-even dichotomy in Heisenberg ladders using the RVB ground states  \cite{gapped2}. It has been noticed that frustrated quantum Heisenberg spin models 
normally possesses short-range dimer states as their ground states. In particular, the ground states of the $J_1- J_2$ model both in 1D and 2D \cite{j1_j2}, the $J_1-J_2-J_3$ AFM Heisenberg model \cite{h2,h3}, and the frustrated AFM 
on the $1/5$-depleted square lattice \cite{h1}, in certain parameter regimes, are the RVB states. Parallely, a family of rotationally invariant spin-$1/2$ Klein Hamiltonians exhibiting ground-state manifolds covered 
by NN valence bond states have also been proposed \cite{h5}. In this direction, a more systematic approach was proposed in which dimer models in different two-dimensional lattices like square, hexagonal, kagom{\'e}, 
are introduced whose exact ground states are valence bond states \cite{dimer model}. 
Further supporting evidence for RVBs being ground states of Heisenberg ladders are provided in 
\cite{theory,h4} and 
\cite{gopal, monte, japan}.
Recent results in the tensor-network formalism reveal that RVB states can be used efficiently to simulate the ground state properties of kagom{\'e} \cite{kag-new} and the $J_1-J_2$ square Heisenberg models \cite{square-new}.

In our work, we assume short-range RVB states, with NN dimer coverings, as the possible ground state of spin-1/2 Heisenberg ladders. 
{Numerical investigation of the ground state from exact diagonalization and the RVB theory, for spin ladders upto 16 spins, provides considerable support for the RVB ansatz from the evaluation of the fidelity ($\cal{F}$) and the normalized relative difference in average energy ($\Delta E$) \cite{comm} between the exact ground states and the RVB states. 
For example, for both the 2-leg and 3-leg quantum spin ladders, upto 16-spins, $\cal{F}$ as high as 0.9 and $\Delta E$ as low as 0.04 are obtained. }
%
%
%
%
{These numerical findings
gives us a good motivation for investigating the genuine multisite entanglement properties of the spin-1/2 Heisenberg ladder using the RVB ansatz.}
Let us also mention here that RVB theory has been also popularized as a possible theoretical tool to understand high-$T_c$ superconductivity \cite{hightc, hightclad} 
and are important in investigating cooperative phenomena in quantum many-body systems \cite{fazio,sir-mam,our_rvb}, 
and related to fault-tolerant quantum computation \cite{fault}.

\section{Resonating valence bond ladders} 
\label{dmrm}
Consider a quantum spin-1/2 ladder, with $\mathcal{L}$ ``legs'' and $\mathcal{M}$ ``rungs'' on a bipartite lattice ($A,B$), comprised of $M (=\mathcal{M})$ sites along the horizontal side and $L (=\mathcal{L})$ sites the vertical side. The total number of spins, $n~ (= M.N)$, is always even, to allow for complete dimer coverings.
Now if the interactions between the spins are restricted to be short-ranged and isotropic, we assume that only NN dimer coverings are allowed. The {equal weight} superposition of all such possible dimer coverings on the lattice would give us  the so-called RVB state, given by 
\begin{equation}
|\psi\rangle = \sum_{\cal{C}}~ \big[|A_1, B_1\rangle \otimes |A_2, B_2\rangle \otimes....|A_N, B_N\rangle \label{eqn:first} \big]_{\cal{C}},
\end{equation} 
where ${\cal{C}}$ refers to a complete dimer covering of the lattice with the summation running over all the coverings, and $|A_i, B_j \rangle$ refers to the dimer $\frac{1}{\sqrt{2}}(|\uparrow_i\rangle |\downarrow_j\rangle -(|\downarrow_i\rangle |\uparrow_j\rangle )$, formed between spins at sites $i$ and $j$, on the sublattices $A$ and $B$, respectively. The RVB state $|\psi \rangle$
is rotationally invariant and is always genuinely multisite entangled state in the {asymptotic limit \cite{prl}.}

{The RVB state in Eq.~(\ref{eqn:first}) is unique. This is done by defining the RVB state on a bipartite lattice ($A,B$). A bipartite lattice is formulated by dividing the spin lattice into two sublattices $A$ and $B$, such that a spin in sublattice $A$ has spins in sublattice $B$ as its nearest neighbours, and vice-versa. In our formalism, we require that all NN dimer states are directed from spins on sublattice $A$ to spins on sublattice $B$, which removes possible ambiguity in the sign of the ground state, and ensures that the defined RVB state is unique.}

Let the RVB state, defined in Eq.~(\ref{eqn:first}), for a quantum spin ladder be denoted by $| {\cal{M}}, {\cal{L}}\rangle$, with $\mathcal{L}$ legs and $\mathcal{M}$ rungs.
Now, let us consider the system containing $M = {\cal{M}}+2$ spins along the rungs, and $L = {\cal{L}}$ number of spins along the legs.  For even $\mathcal{L}$, the state with open boundary condition can be generated recursively as \cite{prl,njp,rvb_more,jpa}
\begin{eqnarray}
|{\cal{M}}+2,{\cal{L}}\rangle&=&|{\cal{M}}+1,{\cal{L}}\rangle|1\rangle_{m+2} +|{\cal{M}},{\cal{L}} \rangle|\bar{2}\rangle _{m+1,m+2}\nonumber \\ 
&=&|{\cal{M}, {\cal{L}}}\rangle|2\rangle_{m+1,m+2} + |{\cal{M}}-1, {\cal{L}}\rangle \nonumber \\
&\times& |\bar{2}\rangle_{m,m+1}|1\rangle_{m+2},
\label{eqn:recur-non_per}
\end{eqnarray}
where $|2\rangle_{m+1,m+2}$  and $|1 \rangle_{m+2}$ correspond to the RVB ladder states, $|2, {\cal{L}} \rangle$ and $|1, {\cal{L}} \rangle$, respectively, and $|\bar{2}\rangle_{m+1,m+2}=|2\rangle_{m+1,m+2}-|1\rangle_{m+1}|1\rangle_{m+2}$. Here, the subscripts denote the rung index. Since, for an $\mathcal{L}$-legged ladder, the index ${\cal{L}}$ is constant in the recursion relations, without loss of generality, we can remove it in the state description, so that the RVB state is denoted by $|{\cal{M}} \rangle$. Incorporation of the periodic boundary condition leads to the following extension of Eq.~(\ref{eqn:recur-non_per}) \cite{njp,prl}:
\begin{eqnarray}
|{\cal{M}}+2\rangle^{{\cal{P}}} =|{\cal{M}}+2\rangle_{1,m+2}+|{\cal{M}}\rangle_{2,m+1}  |\bar{2}\rangle_{m+2,1},
\label{eqn:recur_per}
\end{eqnarray}
where all the terms on the right {can be calculated by using Eq.~(\ref{eqn:recur-non_per}) for RVB states with open boundary condition}. Hence, and hereafter, the superscript $\mathcal{P}$ will indicate that periodic boundary condition has been used for the corresponding state. {Using the recursive relation given in Eq.~(\ref{eqn:recur_per}) we obtain the density matrix characterizing the periodic RVB ladder system, which is given by} 
\begin{eqnarray}
\rho_{({\cal{M}}+2)}^{{\cal{P}}}&=&\rho_{({\cal{M}}+2)}+|{\cal{M}}\rangle \langle {\cal{M}}|_{2,m+1}\otimes|\bar{2}\rangle \langle \bar{2}|_{1,m+2}\nonumber\\
&+&(|{\cal{M}}+2\rangle \langle {\cal{M}}|_{2,m+1} \langle {\bar{2}}|_{m+2,1}+\textrm{h.c}),
\label{neweq}
\end{eqnarray}
{where, the term $\rho_{({\cal{M}}+2)}$ corresponds to the density matrix of the non-periodic RVB ladder}, computed using Eq.~(\ref{eqn:recur-non_per}). 
  
As mentioned earlier, our main interest lies in the multisite entanglement properties of these RVB ladders. In order to explore this, we first need to have expressions {for all possible reduced density matrices of the system. The maximal Schmidt coefficients obtained from these reduced density matrices allow us to compute the GGM of the RVB ladder, using Eq.~(\ref{GGM}). 
As the number of spins in the RVB ladder increases, there is a rapid growth of the number of possible reduced density matrices. However, the symmetry of a periodic RVB state can be exploited to obtain the maximal Schmidt coefficient, which is required to compute GGM in Eq.~(\ref{GGM}), without considering all possible reduced states. For example, extensive numerical studies upto 16 spins confirm that for an $|{\cal{M}}, {\cal{L}}\rangle$ ladder, optimization over the restricted set of all reduced density matrices contained within a reduced $2 \times \cal{L}$ block, say at sites $m+1$ and $m+2$, is sufficient to obtain the maximum Schmidt coefficient for calculating the GGM. The symmetry and periodicity of the RVB ladder ensures that all reduced $2 \times \cal{L}$ block, between any adjacent pair of sites in the lattice, are topologically equivalent.  This reduces the computational difficulty in calculating the genuine multisite entanglement as the optimization over all reduced states is now limited to a $2 \times \cal{L}$ block, which can be analytically derived using a recursion method as discussed in the following segments.}

{For an RVB ladder with open boundary, the reduced density matrix of a $2 \times \cal{L}$ block is obtained by tracing out all the spins except those at the rungs $m+1$ and $m+2$, as given by}
\begin{eqnarray}\label{eqn:NP}
\rho_{(m+1,m+2)}&=&{\cal{N}}_{\cal{M}} |2\rangle \langle 2|_{(m+1,m+2)}+{\cal{N}}_{{\cal{M}}-1} {\bar{\rho}}_{m+1}\nonumber \\
&\otimes&|1 \rangle\langle 1|_{(m+2)} + (|2\rangle_{m+1,m+2} \langle 1|_{m+2} \langle \chi_{{\cal{M}}}|_{m+1}+ \textrm{h.c.})
\end{eqnarray}
where ${\cal{N}_{\cal{M}}}=\langle {\cal{M}} | {\cal{M}}\rangle$ and 
\begin{eqnarray}
\bar{\rho}_{m+1}&=&\textrm{tr}_m(|\bar{2}\rangle \langle \bar{2}|_{m,m+1}),~~~\mathrm{and} \\ 
\langle \chi_{{\cal{M}}}|_{m+1} &=& \langle \bar{2}|_{m,m+1} \langle {\cal{M}}-1| {\cal{M}}\rangle.
\end{eqnarray}

By using Eq.~(\ref{eqn:NP}) we obtain the {reduced density matrix for the $2 \times \cal{L}$ block at rungs $m+1$ and $m+2$, for the periodic RVB ladder state, $\rho_{({\cal{M}}+2)}^{{\cal{P}}}$, given by Eq.~(\ref{neweq}). The reduced density matrix is given by}
\begin{eqnarray}
\rho^{{\cal{P}}}_{(m+1,m+2)}&=&\rho_{m+1,m+2}+\textrm{tr}_{1\cdots m}\big[|{\cal{M}}\rangle \langle {\cal{M}}|_{2,m+1}|\bar{2}\rangle \langle \bar{2}|_{1,m+2}\big]\nonumber\\
&+& ( {|{\cal{M}}\rangle_{2,m+1} |\bar{2}\rangle_{1,m+2} |\langle {\cal{M}}+2|+\textrm{h.c.}}),\nonumber\\
&=&\rho_{m+1,m+2}+\xi^1_{m+1,m+2}+(\xi^2_{m+1,m+2}+\textrm{h.c.}),
\label{even-per}
\end{eqnarray}
where 
\begin{eqnarray}
\xi^1_{m+1,m+2}&=&{\cal{N}}_{{\cal{M}}-1} |1\rangle \langle 1|_{m+1} \otimes \bar{\rho}_{m+2}+{\cal{N}}_{{\cal{M}}-2}  {\bar{\rho}}_{m+1}\nonumber \\
&\otimes& \bar {\rho}_{m+2}+(|1\rangle \langle \chi_{{\cal{M}}-1}|_{m+1}) \otimes \bar{\rho}_{m+2}+h.c.),\nonumber\\\\
\xi^2_{m+1,m+2}&=&|2\rangle _{m+1,m+2} \langle 1|_{m+1}  \langle \chi_{{\cal{M}}}|_{m+2}+|2\rangle_{m+1,m+2} \nonumber\\
&\times& {\bf \sum^{{\cal{M}}}_1} \langle {\cal{K}}_i|_{m+2} \langle \chi_{{\cal{M}}-i}|_{m+1} +\bar{\rho}_{m+1} \otimes|1\rangle_{m+2}\nonumber\\
&\times& \langle \chi_{\cal{M}} |_{m+2}+ 1/{\cal{N}}_1(|{\cal{K}}_1\rangle_{m+1}|1\rangle_{m+2})\nonumber\\
&\times&  \langle 1| _{m+1} {\bf \sum^{{\cal{M}}}_{i=1}} \langle {\cal{K}}|_{m+2} {\cal{J}}^1_{{\cal{M}}-1}.
\label{even-per1}
\end{eqnarray}
Here $ {\cal{J}}^1_{{\cal{M}}}= \langle {\cal{M}} |{\cal{M}}-1\rangle$ and $\langle {\cal{K}}_i|_{m+1}=~_{m,m+1}\langle \bar{2}|{\cal{K}}_{i-1}\rangle_{m}$ with $|{\cal{K}}_0\rangle_m=|1\rangle_m$. The recursion relation for the inner product $\langle {\cal{M}}|{\cal{M}}\rangle$ can now be expressed as 
\begin{eqnarray}
{\cal{N}}_{{\cal{M}}}&=& {\cal{N}}_1 {\cal{N}}_{{\cal{M}}-1}+{\cal{N}}'_2 {\cal{N}}_{{\cal{M}}-2}+2(-1)^{m-1} {\bf \sum _i}  {\gamma^1_i}{\cal{J}}_{{\cal{M}}-1},
\end{eqnarray}
where ${\cal{N}}'_2=\langle \bar{2}|\bar{2}\rangle$ and all the $\gamma_i$'s can be calculated using the linear equation 
\begin{equation}
\langle \gamma_i|\bar{2}\rangle_{m,m+1}=(-1)^{m-1}{\bf{\sum}}^{j_2}_{j=j1} \gamma_j| \gamma_j\rangle_{m+1},
\end{equation}
where $|\gamma_j\rangle$ form an independent set of vectors consisting of certain singlet combinations of an $(1, \mathcal{N}+2)$ spin system, e.g $|\gamma_1\rangle= |1\rangle$.  
  
For odd $\cal{L}$, the recursion relations are much simpler as the number of possible coverings is lower. The recursion relation for the RVB ladder with periodic boundary conditions is given by
\begin{eqnarray}
|{\cal{M}}+2,{\cal{L}}\rangle_{\cal{P}}&=&|{\cal{M}},{\cal{L}}\rangle_{1,m}|2\rangle_{m+1,m+2} \nonumber \\ 
&+&|{\cal{M}},{\cal{L}} \rangle_{2,m+1}|2\rangle_{m+2,1}.
\end{eqnarray}
The reduced density matrix {for the $2 \times \cal{L}$ block, at sites $m+1$ and $m+2$,} corresponding to the above state is given by
\begin{eqnarray}
{\rho^{{\cal{P}}}_{(m+1,m+2)}}&=&\mathcal{N}_{\cal{M}}|2\rangle\langle2|_{m+1,m+2} + \mathcal{N}_{\mathcal{M}-2} \rho_{m+1} \otimes \rho_{m+2}\nonumber\\
&+& \left(|2\rangle_{m+1,m+2}\langle\xi^3|_{m+1,m+2} + \textrm{h.c.}\right),
\label{odd-per}
\end{eqnarray}
where 
\begin{eqnarray}
\langle\xi^3|_{m+1,m+2} &=& \langle2|_{1,m+2}\langle\mathcal{M}|_{2,m+1}|\mathcal{M}\rangle_{1,m},~~\textrm{and}\\
\rho_{m+1}&=& \textrm{tr}_m(|2\rangle\langle2|_{m,m+1})
\label{odd-per1}
\end{eqnarray} 
Hence, using Eqs.~(\ref{even-per}) and (\ref{odd-per}) for even- and odd-legged RVB ladders, respectively, one can obtain the reduced density matrices {for the $2 \times \cal{L}$ block} necessary to compute the generalized geometric measure. {We note that the maximal Schmidt coefficient is obtained by considering the reduced states within the $2 \times \cal{L}$ block. It is observed that the maximum Schmidt coefficients are typically obtained from the $2:\textrm{rest}$ or the $4:\textrm{rest}$ bipartitions where the reduced spins are nearest neighbors, though there does not seem to be any distinctive pattern that can systematically differentiate between the typical bipartitions in odd and even ladders. Moreover, no systematic pattern is observed in which topologically inequivalent reduced states provide the same maximum Schmidt coefficient.}  

We subsequently compare the GGM of the RVB state with that of the ground state of the Heisenberg ladder obtained by exact diagonalization. Note here that although we use the above method for calculating the GGM, the same recursion can be used to calculate other system properties like magnetization, susceptibility, classical correlators, bipartite entanglement and other quantum correlations, etc. A more developed exposition and formalism for the density matrix recursion method can be obtained in Refs. \cite{njp,prl}.

\section{Diverging scaling with converging multisite entanglement}
\label{ggm}
Applying the recursion technique, we can investigate the behavior of genuine multisite entanglement of the RVB state in large quantum spin lattices. For example, one can study the finite-size scaling of GGM in $\mathcal{L}$-legged ladders with large number of rungs and investigate the odd-even dichotomy in the asymptotic limit. 
For odd-legged RVB ladders, the GGM initially increases with increasing number of even rungs, $\mathcal{M}$, before approaching a constant value at large $\mathcal{M}$ (see Fig.~{\ref{fig:2}}(b)), while for even-legged ladders, the GGM decreases with increasing $\mathcal{M}$, before flattening to a constant for larger number of rungs as shown in the inset of Fig.~\ref{fig:3}. Note that the DMRM approach is not possible to access an odd number of rungs. Importantly, we find that the behavior of genuine multisite entanglement of the ground state of the Heisenberg ladder with even-  and odd-legged ladders is qualitatively similar to results obtained with the RVB ansatz. This is clearly seen by comparing Fig.~\ref {fig:2}(a) with \ref{fig:2}(b), and the main with inset in Fig.~\ref{fig:3}.

 
The similarity between the two methods, viz. exact diagonalization of the Heisenberg ladder and RVB ansatz on the same lattice, motivates us to perform finite-size scaling analysis of GGM, by using RVB theory, wherein we can handle large lattice sizes. The analysis would shed light on the finite-size behavior of multiparty entanglement of the original Heisenberg ladder. 
%
The finite-size scaling of GGM in a pure quantum spin ground state, $|\psi\rangle$, can be analyzed through the scaling relation, 
$
\mathcal{G}(|\psi\rangle) \approx \mathcal{G}_c(|\psi\rangle) \pm k n^{-x},
$
where, $n$ is the total number of spins, $\mathcal{G}_c$ is an estimated value of GGM at high $n$, 
and $x$ is the ``scaling exponent'' with which the GGM approaches its asymptote at large $n$. 
For an $\mathcal{L}$-legged RVB ladder, written as $|\mathcal{L},\mathcal{M}\rangle$, the finite-size scaling is given by the relation,
\begin{equation}
\label{exponent}
\mathcal{G}(|\mathcal{L},\mathcal{M}\rangle) \approx \mathcal{G}_c(\mathcal{L}) \pm k n^{-x(\mathcal{L})}.
\end{equation}
Using the DMRM method, we have computed the GGM for RVB ladders upto $\mathcal{L} < 8$, with $\mathcal{M} =$ 20.  Once can easily extend the computation for higher values of $\mathcal{M}$. 

\begin{figure}[h]
\begin{center}
\epsfig{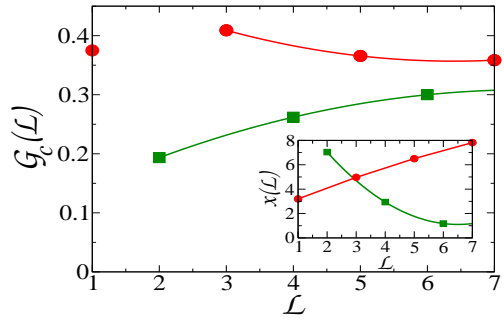}
\caption{Diverging scaling with converging multisite entanglement. The behavior of the asymptotic GGM ($\mathcal{G}_c(\mathcal{L})$) and the scaling exponent ($x(\mathcal{L})$) with increasing $\mathcal{L}$. We observe that even though,  the $\mathcal{G}_c(\mathcal{L})$  for odd- and even-legged ladders, converge with increasing $\mathcal{L}$, the scaling $x(\mathcal{L})$ diverge. All axes are dimensionless.}\label{fig:1}
\end{center}
\end{figure}

Fig.~\ref{fig:1} shows the values of $\mathcal{G}_c(\mathcal{L})$ and $x(\mathcal{L})$, for different values of $\mathcal{L}$, where the GGM is scaled upto $\mathcal{M}$ = 20 rungs for an RVB ladder of $\mathcal{L}$ legs. We observe that as $\mathcal{L}$  increases, the $\mathcal{G}_c(\mathcal{L})$ for the odd- ladders converges to that for the even ones. This is consistent since the pseudo-2D spin ladders slowly approaches the square-2D lattice, and in the asymptotic limit, one cannot distinguish whether the system was originally generated by increasing $\mathcal{L}$ in an odd- or even-legged ladder.  However, we find that the scaling exponent, $x(\mathcal{L})$, for odd and even ladders, converges to \emph{ different} values with increase of $\mathcal{L}$ (see inset of Fig.~\ref{fig:1}). We therefore have a diverging scaling exponent 
for odd- and even-legged ladders, even though the corresponding multisite entanglement converge. The diverging $x(\mathcal{L})$, therefore, shows that the finite-size scaling of GGM for 
RVB ladders can highlight the odd-even dichotomy at large $\mathcal{L}$.
The results show that the GGM for odd RVB ladders converges slower than that for even ladders at low $\mathcal{L}$, which is reversed as $\mathcal{L}$ is increased and holds even at large $\mathcal{L}$, where $\mathcal{G}_c(\mathcal{L})$ for odd and even RVB ladders are indistinguishable.

Therefore, we observe that although the value of genuine multisite entanglement can not distinguish the odd-legged ladders from the even-legged ones for large lattice size, the corresponding finite-size scaling exponents are capable of detecting the difference.

\section{Discussion}
\label{conclude}
To summarize, we investigate the behavior of genuine multipartite entanglement of the ground state in odd- and even- Heisenberg ladders. Even though such models have immense fundamental and
practical importance, owing in particular to the dissimilarities on the two sides of the odd-versus-even divide, they are not analytically accessible. In this work, we began our investigation through exact diagonalization techniques
to  find that the genuine multisite entanglement, as quantified by the GGM, of the ground state obtained from the odd-legged ladder, increases with the number of rungs. The opposite is true 
in the even-legged ladder.
This feature is in good qualitative agreement with the assumption that ground states of odd and even Heisenberg ladders are RVB states.
We perform scaling analyses of the RVB states on ladders of large system sizes by employing the DMRM, and find that 
while the GGM of the RVB states on large ladders converges to a single value independent of the odd-even parity of the ladders, their 
scaling exponents diverge from each other.
%
%
%
%
While the study reported is for the isotropic Heisenberg model, we have carried out parallel studies for the quantum $XXZ$ model. We observed qualitative similarity of the results obtained for values of the $zz$ vs. $xx$ anisotropy up to approximately 1.4.


\section*{References}
\begin{thebibliography}{10}








\bibitem{many}
S. R. White, R. M. Noack, and D. J. Scalapino, 
\emph{Resonating Valence Bond Theory of Coupled Heisenberg Chains},
Phys. Rev. Lett. \textbf{73}, 886 (1994).

\bibitem{daggoto}
E. Dagotto, and T. M. Rice, 
\emph{Surprises on the Way from 1D to 2D Quantum Magnets: the Novel Ladder Materials}, 
 Science {\bf 271}, 618  (1996).  


\bibitem{hightclad} E. Dagotto, J. Riera, and D. Scalapino, 
\emph{Superconductivity in ladders and coupled planes},
Phys. Rev. B \textbf{45}, 5744 (1992).



%

%










\bibitem{gopal}S. Gopalan, T. M. Rice, and M. Sigrist,
\emph{Spin ladders with spin gaps: A description of a class of cuprates},
Phys. Rev. B \textbf{49}, 8901 (1994).


\bibitem{gapped}T. M. Rice, S. Gopalan, and M. Sigrist,
\emph{Superconductivity, Spin Gaps and Luttinger Liquids in a Class of Cuprates},
Europhys. Lett. \textbf{23}, 445 (1993).


\bibitem{entanglement} R. Horodecki, P. Horodecki, M. Horodecki, and K. Horodecki, 
\emph{Quantum entanglement},
Rev. Mod. Phys. {\bf 81}, 865 (2009).

\bibitem{modi}
K. Modi, A. Brodutch, H. Cable, T. Paterek, and V. Vedral,
\emph{The classical-quantum boundary for correlations: Discord and related measures},
Rev. Mod. Phys. {\bf 84}, 1655 (2012).
 
\bibitem{sachdev} S. Sachdev,
\emph{Quantum Phase Transitions}
(Cambridge Univ. Press, Cambridge, 1999).

\bibitem{fazio}
L. Amico, R. Fazio, A. Osterloh, and V. Vedral,
\emph{Entanglement in many-body systems}, 
Rev. Mod. Phys. {\bf 80}, 517 (2008).

\bibitem{sir-mam}
M. Lewenstein, A. Sanpera, V. Ahufinger, B. Damskic, A. Sen(De), and U. Sen,
\emph{Ultracold atomic gases in optical lattices: mimicking condensed matter physics and beyond},
Adv. Phys. {\bf{56}}, 243 (2007).

\bibitem{kit} A. Kitaev and J. Preskill, 
\emph{Topological Entanglement Entropy},
Phys. Rev. Lett. {\bf 96}, 110404 (2006).

\bibitem{wen} M. Levin and X.-G. Wen,
\emph{Detecting Topological Order in a Ground State Wave Function},
Phys. Rev. Lett. {\bf 96}, 110405 (2006).

\bibitem{wen2} X. Chen, Z.-C. Gu, and X.-G. Wen,
\emph{Local unitary transformation, long-range quantum entanglement, wave function renormalization, and topological order},
Phys. Rev. B \textbf{82}, 155138 (2010).


\bibitem{multi_others}
T.-C. Wei, D. Das, S. Mukhopadyay, S. Vishveshwara,
and P. M. Goldbart, 
\emph{Global entanglement and quantum criticality in spin chains},
Phys. Rev. A {\bf 71}, 060305(R) (2005); 
T. R. de Oliveira, G. Rigolin, M. C. de Oliveira,
and E. Miranda, 
\emph{Multipartite Entanglement Signature of Quantum Phase Transitions},
Phys. Rev. Lett. {\bf 97}, 170401 (2006);
D. Buhr, M. E. Carrington, T. Fugleberg, R. Kobes, G.
Kunstatter, D. McGillis, C. Pugh, and D. Ryckman, 
\emph{Geometrical entanglement of highly symmetric multipartite states and the Schmidt decomposition},
J. Phys. A: Math. Theor. {\bf 44}, 365305 (2011).



\bibitem{multiparty}
M.-F. Yang, 
\emph{Reexamination of entanglement and the quantum phase transition},
Phys. Rev. A {\bf71}, 030302 (2005);
X.-F. Qian, T. Shi, Y. Li, Z. Song, and C.-P. Sun, 
\emph{Characterizing entanglement by momentum jump in the frustrated Heisenberg ring at a quantum phase transition}, Phys.
Rev. A {\bf 72}, 012333 (2005);
M. N. Bera, R. Prabhu, A. Sen(De), and U. Sen,
\emph{Multisite Entanglement acts as a Better Indicator of Quantum Phase Transitions in Spin Models with Three-spin Interactions},
arXiv:1209.1523;
A. Biswas, R. Prabhu, A. Sen(De), and U. Sen,
\emph{Genuine-multipartite-entanglement trends in gapless-to-gapped transitions of quantum spin systems},
Phys. Rev. A {\bf 90}, 032301 (2014).


\bibitem{hast}A. B. Kallin, I. Gonz{\'a}lez, M. B. Hastings, and R. G. Melko,
\emph{Valence Bond and von Neumann Entanglement Entropy in Heisenberg Ladders},
Phys. Rev. Lett. \textbf{103}, 117203 (2009);
%
H. Ju, A. B. Kallin, P. Fendley, M. B. Hastings, and R. G. Melko,
\emph{Entanglement scaling in two-dimensional gapless systems},
Phys. Rev. B \textbf{85}, 165121 (2012).
  
\bibitem{ren} M. B. Hastings, I. Gonz{\'a}lez, A. B. Kallin, and R. G. Melko,
\emph{Measuring Renyi Entanglement Entropy in Quantum Monte Carlo Simulations},
Phys. Rev. Lett. \textbf{104}, 157201 (2010).

\bibitem{njp}
H. S. Dhar, A. Sen(De), and U. Sen, 
\emph{The density matrix recursion method: genuine multisite entanglement distinguishes odd from even
quantum spin ladder states},
New J. Phys. {\bf 15}, 013043 (2013).


\bibitem{ggm}A. Sen(De) and U. Sen,
\emph{Channel capacities versus entanglement measures in multiparty quantum states},
Phys. Rev. A {\bf 81}, 012308 (2010);
A. Sen(De) and U. Sen,
\emph{Bound Genuine Multisite Entanglement: Detector of Gapless-Gapped Quantum Transitions in Frustrated Systems},
arXiv:1002.1253 (2010).

\bibitem{geometric}
A. Shimony,
\emph{Degree of entanglement},
Ann. NY Acad. Sci. {\bf755}, 675 (1995); 
H. Barnum
and N. Linden,\emph{ Monotones and invariants for multi-particle quantum states}, J. Phys. A {\bf 34}, 6787 (2001);
T. Wei and P. M. Goldbart,
\emph{Geometric measure of entanglement and applications to bipartite and multipartite quantum states},
Phys. Rev. A {\bf 68}, 042307 (2003);
R. Or{\'u}s, \emph{Universal geometric entanglement close to quantum phase transitions},
Phys. Rev. Lett. {\bf100}, 130502 (2008); 
R. Or{\'u}s, S. Dusuel, and J. Vidal, 
\emph{Equivalence of critical scaling laws for many-body entanglement in the Lipkin-Meshkov-Glick model},
ibid. {\bf101}, 025701 (2008); 
R. Or{\'u}s,  
\emph{Geometric entanglement in a one-dimensional valence-bond solid }, 
Phys. Rev. A {\bf 78}, 062332 (2008);
Q.-Q. Shi, R. Or{\'u}s, J. O. Fj{\ae}restad, and H.-Q. Zhou,
\emph{Finite-size geometric entanglement from tensor network algorithms},
 New J. Phys. {\bf 12}, 025008 (2010); 
 R. Or{\'u}s and T.-C. Wei, 
 \emph{Visualizing elusive phase transitions with geometric entanglement}, 
 Phys. Rev. B {\bf 82}, 155120 (2010);
M. Blasone, F. Dell’Anno, S. DeSiena, and F. Illuminati,
\emph{Hierarchies of geometric entanglement}, 
 Phys. Rev. A {\bf 77}, 062304 (2008).

\bibitem{lanc}C. Lanczos, 
\emph{An Iteration Method for the Solution of the Eigenvalue Problem of Linear Differential and Integral Operators},
J. Res. Natl. Bur. Stand. {\bf 45}, 255-282 (1950).

\bibitem{tit} H. Nishimori, 
\emph{Titpack--Numerical diagonalization routines and quantum spin Hamiltonians},
AIP Conf. Proc. \textbf{248}, 269 (1991).


\bibitem{rvb_more}
G.  Sierra and M. Delgado,
 \emph{Short-range resonating-valence-bond state of even-spin ladders: A recurrent variational approach}, M A  Phys. Rev. B {\bf 56} 8774 (1997);
 M. Roncaglia, G. Sierra, and M. A. Martin-Delgado, 
 \emph{Dimer–resonating valence bond state of the four-leg Heisenberg ladder: Interference among resonances},
 Phys. Rev. B {\bf 60}, 12134 (1999).

\bibitem{prl} H. S. Dhar, A. Sen(De), and U. Sen, 
\emph{Characterizing Genuine Multisite Entanglement in Isotropic Spin Lattices},
Phys. Rev. Lett. {\bf 111}, 070501 (2013).




\bibitem{comment} The fundamental obstacle in computing the genuine multisite entanglement for states of large quantum systems is essentially two-fold: The inability to obtain accurate eigenstates of system Hamiltonians, by exact diagonalization and insufficient computational power to compute the reduced density matrices across all bipartition.



\bibitem{chiral}  E. Orignac and T. Giamarchi, 
\textit{Meissner effect in a bosonic ladder},
Phys. Rev. B \textbf{64}, 144515 (2001);
%
M. Atala, M. Aidelsburger, M. Lohse, J. T. Barreiro, B. Paredes, and I. Bloch,
\textit{Observation of chiral currents with ultracold atoms in bosonic ladders},
Nat. Phys. \textbf{10}, 588 (2014);    
%
A. Petrescu and K. L. Hur,
\emph{Chiral Mott insulators, Meissner effect, and Laughlin states in quantum ladders},
Phys. Rev. B {\bf 91}, 054520 (2015).


\bibitem{optical} I. Bloch, 
\emph{Exploring quantum matter with ultracold atoms in optical lattices},
J. Phys. B: At. Mol. Opt. Phys. \textbf{38}, S629 (2005).


\bibitem{op-ion}
P. Treutlein, T. Steinmetz, Y. Colombe, B. Lev, P. Hommelhoff,
J. Reichel, M. Greiner, O. Mandel, A. Widera,
T. Rom, I. Bloch, and T. W. H¨ansch, 
\emph{Quantum information processing in optical lattices and magnetic microtraps},
Fortschr. Phys. {\bf 54},
702 (2006).

\bibitem{ion}
D. Leibfried, R. Blatt, C. Monroe, and D. Wineland,
\emph{Quantum dynamics of single trapped ions},
Rev. Mod. Phys. \textbf{75}, 281 (2003); 
H. H{\"a}ffner, C. Roos, and R.Blatt, 
\emph{Quantum computing with trapped ions},
Phys. Rep. \textbf{469}, 155 (2008);
K. Ki, M.-S. Chang,	S. Korenblit,	R. Islam, E. E. Edwards, J. K. Freericks, G.-D. Lin, L.-M. Du, and C. Monroe,
\emph{Quantum simulation of frustrated Ising spins with trapped ions},
Nature {\bf 465}, 590 (2010).

\bibitem{photon}
A. Aspuru-Guzik and P. Walther,
\emph{Photonic quantum simulators},
Nat. Phys. {\bf8}, 285 (2012).

\bibitem{rvb_photon} X.-S. Ma, B. Dakic, W. Naylor, A. Zeilinger, and P. Walther, 
\emph{Quantum simulation of the wavefunction to probe frustrated Heisenberg spin systems},
Nat. Phys. \textbf{7}, 399 (2011).




\bibitem{nmr}L. M. K. Vandersypen and I. L. Chuang,
\emph{NMR techniques for quantum control and computation},
Rev. Mod. Phys. \textbf{76}, 1037 (2005).


\bibitem{rvb_optical}S. Nascimb{\'e}ne, Y. A. Chen, M. Atala, M. Aidelsburger, S. Trotzky, B. Paredes, and I. Bloch, 
\emph{Experimental Realization of Plaquette Resonating Valence-Bond States with Ultracold Atoms in Optical Superlattices},
Phys. Rev. Lett. \textbf{108}, 205301 (2012).



\bibitem  {bethe-anstaz} 
H.A. Bethe,
\emph{On the theory of metals.}
 Z.Phys. {\textbf 71}, 205 (1931).

\bibitem{dmrg} S.R. White, 
\emph{Density matrix formulation for quantum renormalization groups},
Phys. Rev. Lett. \textbf{69}, 2863 (1992); 
\emph{Density-matrix algorithms for quantum renormalization groups}
Phys. Rev. B \textbf{48}, 10345 (1993); 
U. Schollw{\"o}ck, 
\emph{The density-matrix renormalization group},
Rev. Mod. Phys. \textbf{77}, 259 (2005).

\bibitem{qmc}W. M. C. Foulkes, L. Mitas, R. J. Needs, and G. Rajagopal,
\emph{Quantum Monte Carlo simulations of solids},
Rev. Mod. Phys. \textbf{73}, 33 (2001).

\bibitem{rvb1}P. W. Anderson, 
\emph{Resonating valence bonds: A new kind of insulator?},
Mater. Res. Bull. {\bf 8}, 153 (1973).

\bibitem{rvb2} S. A. Kivelson, D. S. Rokhsar, and J. P. Sethna,
\emph{Topology of the resonating valence-bond state: Solitons and high-Tc superconductivity},
Phys. Rev. B {\bf 35}, 8865 (1987);
S. Liang, B. Doucot, and P. W. Anderson,
\emph{Some New Variational Resonating-Valence-Bond-Type Wave Functions for the Spin-½ Antiferromagnetic Heisenberg Model on a Square Lattice},
Phys. Rev. Lett.{\bf 61}, 365 (1988);
G. Misguich, D. Serban, and V. Pasquier,
 \emph{Quantum Dimer Model on the kagom{\'e} Lattice: Solvable Dimer-Liquid and Ising Gauge Theory},
Phys. Rev. Lett.{\bf 89}, 137202 (2002).


\bibitem{gapped2} H. Yao and S. A. Kivelson,
\emph{Exact Spin Liquid Ground States of the Quantum Dimer Model on the Square and Honeycomb Lattices},
Phys. Rev. Lett. \textbf{108}, 247206 (2012).


\bibitem{j1_j2}
C. K. Majumdar and D. K. Ghosh,
\emph{On Next-Nearest-Neighbor Interaction in Linear Chain},
J. Math. Phys. {\bf 10}, 1388 (1969);
B. S. Shastry and B. Sutherland,
{ \emph Exact Ground State of a Quamtum Mechanical antiferromagnet}, 
Physica { \bf108B}, 1069 (1981);
H. J. Mikeska and A. K. Kolezhuk,
\emph{One-dimensional magnetism in: Quantum Magnetism}, eds. U.  Schollw\"ock, J. Richter, D.J.J. Farnell
and R.F. Bishop, Lecture Notes in Physics, {\bf 645} (Springer, Berlin, 2004).

\bibitem{h3}
M. Mambrini, A. Lauchli, D. Poilblanc, and F. Mila, 
\emph{Plaquette valence-bond crystal in the frustrated Heisenberg quantum antiferromagnet on the square lattice},
Phys. Rev. B \textbf{74}, 144422 (2006).

\bibitem{h2} B. Kumar, 
\emph{Quantum spin models with exact dimer ground states},
Phys. Rev. B \textbf{66}, 024406 (2002).

\bibitem{h1} I. Bose and A. Ghosh,
\emph{Exact ground and excited states of frustrated antiferromagnets on the $CaV_4O_9$ lattice},
Phys. Rev. B \textbf{56}, 3149 (1997).

\bibitem{h5} K. S. Raman, R. Moessner, and S. L. Sondhi,
\emph{$SU(2)$-invariant spin-$\frac{1}{2}$ Hamiltonians with resonating and other valence bond phases},
Phys. Rev. B \textbf{72}, 064413 (2005).

\bibitem{dimermodel} R. Moessner, S. L. Sondhi, and P. Chandra,
\emph{Phase diagram of the hexagonal lattice quantum dimer model},
Phys. Rev. B {\bf 64}, 144416 (2001); R. Moessner and S. L. Sondhi,
\emph{Resonating Valence Bond Phase in the Triangular Lattice Quantum Dimer Model},
Phys. Rev. Lett. {\bf 86}, 1881 (2001); 
G. Misguich, D. Serban, and V. Pasquier, \emph{Quantum Dimer Model on the kagom{\'e} Lattice: Solvable Dimer-Liquid and Ising Gauge Theory}, Phys. Rev. Lett. \textbf{89}, 137202 (2002);
G. Misguich and C. Lhuillier, 
\emph{Frustrated spin system} ( World Scientific, Singapore, 2004).

\bibitem{theory}D. S. Rokhsar, and S. A. Kivelson, 
\emph{Superconductivity and the Quantum Hard-Core Dimer Gas},
Phys. Rev. Lett. \textbf{61}, 2376 (1988).





%






\bibitem{h4}
J. Cano and P. Fendley, 
\emph{Spin Hamiltonians with Resonating-Valence-Bond Ground States},
Phys. Rev. Lett. \textbf{105}, 067205 (2010).




\bibitem{monte}B. Frischmuth, B. Ammon, and M. Troyer,
\emph{Susceptibility and low-temperature thermodynamics of spin--$\frac{1}{2}$ Heisenberg ladders},
Phys. Rev. B \textbf{54}, R3714(R) (1996).

\bibitem{japan} Y. Nishiyama, N. Hatano, and M. Suzuki,
\emph{Hidden Orders and RVB Formation of the Four-Leg Heisenberg Ladder Model},
J. Phys. Soc. Japan \textbf{65}, 560 (1996).


\bibitem{kag-new} S. Yan, D.A. Huse, and S.R. White, \emph{Spin-Liquid Ground State of the $S= 1/2$ kagom{\'e} Heisenberg Antiferromagnet}, Science \textbf{332}, 1173 (2011); 
S. Depenbrock, I.P. McCulloch, and U. Schollw{\"o}ck, \emph{Nature of the Spin-Liquid Ground State of the $S=1/2$ Heisenberg Model on the kagom{\'e} Lattice}, Phys. Rev. Lett. \textbf{109}, 067201 (2012); D. Poilblanc, N. Schuch, D. Perez-Garcia, and J.I. Cirac, \emph{Resonating valence bond states in the PEPS formalism}, Phys. Rev. B \textbf{86}, 014404 (2012).

\bibitem{square-new} H.-C. Jiang, H. Yao, and L. Balents,
\emph{Spin liquid ground state of the spin-$\frac{1}{2}$ square $J_1-J_2$ Heisenberg model},
Phys. Rev. B \textbf{86}, 024424 (2012);
L. Wang, Z.-C. Gu, F. Verstraete, and X.-G. Wen, \emph{Spin-liquid phase in spin-$\frac{1}{2}$ square $J_1-J_2$ Heisenberg model: A tensor product state approach}, arXiv:1112.3331; 
L. Wang, D. Poilblanc, Z.-C. Gu, X.-G. Wen, and F. Verstraete,
\textit{Constructing a Gapless Spin-Liquid State for the Spin-$\frac{1}{2}$ $J_1-J_2$ Heisenberg Model on a Square Lattice}, Phys. Rev. Lett. \textbf{111}, 037202 (2013).


\bibitem{comm} If we consider $|\phi_g\rangle$ as the exact ground state of the Hamiltonian, $H_{int}$, and $|\psi\rangle$ as the RVB state, corresponding to a quantum spin-1/2 ladder, then the fidelity between the two states is defined as $|\langle\phi_g|\psi\rangle|$. Further, the normalized relative difference in average energy is obtained using the relation, $\frac{ E_g - \langle\psi|H_{int}|\psi\rangle}{E_g}$, where $E_g = \langle\phi_g|H_{int}|\phi_g\rangle$.

\bibitem{hightc} P. W. Anderson, 
\emph{The Resonating Valence Bond State in $La_2CuO_4$ and Superconductivity},
Science \textbf{235}, 1196 (1987); 


\bibitem{our_rvb}
A. Chandran, D. Kaszlikowski, A. Sen(De), U. Sen, and V. Vedral,
\emph{Regional Versus Global Entanglement in Resonating-Valence-Bond States},
Phys. Rev. Lett. \textbf{99}, 170502 (2007);
H. S. Dhar and A. Sen(De),
\emph{Entanglement in resonating valence bond states: ladder versus isotropic lattices},
J. Phys. A: Math. Theor. \textbf{44} 465302 (2011).


\bibitem{fault} A. Y. Kitaev, 
\emph{Fault-tolerant quantum computation by anyons},
Ann. Phys.\textbf{303}, 2 (2003).


\bibitem{jpa}
 Y. Fan and M. Ma, \emph{Generating-function approach to the resonating-bond state on the triangular and square ladders},
Phys. Rev. B {\bf37}, 1820 (1988).
















\end{thebibliography}
 \end{document}